\documentclass[12pt]{iopart}
\usepackage{iopams}
\usepackage{graphicx}
\usepackage[breaklinks=true,colorlinks=true,linkcolor=blue,urlcolor=blue,citecolor=blue]{hyperref}
\usepackage{amssymb}
\usepackage{rotating}
\begin{document}

\title[Negative thermal expansion in Y$_2$W$_3$O$_{12}$]{Simulation study of negative thermal expansion in yttrium tungstate Y$_2$W$_3$O$_{12}$}

\author{Leila H N Rimmer$^{1,2,3}$ and Martin T Dove$^{1,3}$}

\address{$^1$ School of Physics and Astronomy \textit{and} Materials Research Institute, Queen Mary University of London, Mile End Road, London, E1 4NS, United Kingdom.}
\address{$^2$ CrystalMaker Software Ltd, Centre for Innovation \& Enterprise, Oxford University Begbroke Science Park, Woodstock Road, Begbroke, Oxfordshire, OX5 1PF, United Kingdom.}
\address{$^3$ Department of Earth Sciences, University of Cambridge, Downing Street, Cambridge, CB2 3EQ, United Kingdom.}
\ead{martin.dove@qmul.ac.uk}

\begin{abstract}
A simulation study of negative thermal expansion in Y$_2$W$_3$O$_{12}$ was carried out using calculations of phonon dispersion curves through the application of density functional perturbation theory. The mode eigenvectors were mapped onto flexibility models and results compared with calculations of the mode Gr\"{u}neisen parameters. It was found that many lower-frequency phonons contribute to negative thermal expansion in Y$_2$W$_3$O$_{12}$, all of which can be described in terms of rotations of effectively rigid WO$_4$ tetrahedra and Y--O rods. The results are strikingly different from previous phonon studies of higher-symmetry materials that show negative thermal expansion.
\end{abstract}

\submitto{\JPCM}

\section{Introduction}

\subsection{Atomic-scale origin of negative thermal expansion}

As has been highlighted in several recent reviews \cite{Barrera_2005_review,Romao_2013_NTEReview,Lind_2012_Mater_NTEReview,Takenaka_2012_NTEReview,Miller_2009_NTEReview}, a growing number of materials have been identified as exhibiting \textit{negative thermal expansion} (NTE): the phenomenon whereby a material shrinks, rather than expands, on heating. NTE is found in several framework material families, such as those represented by SiO$_2$ \cite{NTE_quartz_data,NTE_quartz_theory}, Cu$_2$O \cite{Cu2O_Tiano,Cu2O_Sanson,Cu2O_Gupta,Leila_Cu2O}, ScF$_3$ \cite{ScF3_discovery,ScF3_phonons}, ZrW$_2$O$_8$ \cite{ZrW2O8_Science_1996,Pryde_1996,ZrW2O8_structure_1999,Cao_etal_2002_PRL_ZrW2O8NTEFrustratedModes,Cao_etal_2003_PRB_ZrW2O8NTEXAFS}, Sr$_2$W$_3$O$_{12}$ \cite{Sc2W3O12_Evans}, Zn(CN)$_2$ \cite{Goodwin_Zn(CN)2,Hong_Zn(CN)2} as well as in a number of hybrid metal-organic framework materials \cite{MOFs_Han_2007,MOFs_Dubbeldam_2007,MOFs_Wu_2008}. 

Our understanding of positive thermal expansion is based on the nature of interatomic bonding. The potential energy between a pair of bonded atoms is anharmonic and asymmetric about the position of the energy minimum, with the slope of the potential energy function being shallower in the direction of greater interatomic separation. As a result, when the material is heated and the energy of the atoms increases, the average separation of each bonded pair also increases. This implies an overall expansion of the bulk material on heating; that is \textit{positive} thermal expansion, making \textit{negative} thermal expansion a counter-intuitive and unexpected phenomenon.


Our qualitative understanding of NTE in framework materials is based on the fact that, in NTE systems, the energy cost of stretching interatomic bonds is typically much higher than transverse vibrations of those same bonds. For example, we could label two cations \textbf{M} and \textbf{M$^\prime$} that form a \textbf{M}--O--\textbf{M$^\prime$} linkage via a bridging oxygen atom. If the \textbf{M}--O and \textbf{M$^\prime$}--O bonds are very stiff, transverse displacements of the O atom will pull the \textbf{M} and \textbf{M$^\prime$} cations towards each other. Since the amplitude of a transverse displacement will increase on heating, this leads to negative thermal expansion. This mechanism is called the \textit{tension effect} \cite{Barrera_2005_review}. 

The drive towards positive thermal expansion exists in all materials, including those with overall NTE. Therefore, for a material to show NTE, the amplitude of the oxygen displacement must be sufficiently large to outweigh those mechanisms that drive positive thermal expansion. Given that the amplitude of a phonon is inversely proportional to the square of its frequency, an NTE-driving phonon must also have low frequency to achieve this condition. 

All of the NTE-exhibiting material families mentioned above share one significant feature: their crystal structures can all be described in terms of an infinite three-dimensional network of corner-sharing structural polyhedra (such as SiO$_4$, WO$_4$, OCu$_4$ and ZnN$_4$ tetrahedra, and ScF$_6$, ZrO$_6$, and YO$_6$ octahedra). Deformation of coordination polyhedra is likely to require higher energies due to the electrostatic repulsion between oxygens at the polyhedral vertices or to the high sensitivity of d-orbital bonding around a metal coordination centre with respect to bonding geometry. Therefore, any tension effect that avoids or minimises deformation of those polyhedra is more likely to be acting at a low enough frequency to drive macroscopic NTE. 

The conditions outlined above describe a \textit{Rigid Unit Mode} (RUM) \cite{Pryde_1996,Giddy_RUMs,Hammonds_RUMs,Dove_RUMs}: a low frequency phonon with rotations and translations of effectively rigid coordination polyhedra. The RUM model can be used to explain why structures with stiff coordination polyhedra have tension effect vibrations with sufficiently large amplitude to drive NTE \cite{NTE_theory_Welche}. In cases such as ScF$_3$ \cite{ScF3_phonons}, Zn(CN)$_2$ \cite{Hong_Zn(CN)2} and Cu$_2$O \cite{Leila_Cu2O}, lattice dynamics calculations have shown that the phonons responsible for NTE are RUMs involving motion of ScF$_6$ octahedra, Zn(C/N)$_4$ tetrahedra and CuO$_2$ rods respectively moving as rigid units. The case of ZrW$_2$O$_8$ is one that we will consider in a separate publication but, as a prelude to this, we consider here the case of Y$_2$W$_3$O$_{12}$---a material which, intriguingly, \textit{cannot} support RUMs \cite{Hammonds_octahedra} but \textit{does} exhibit NTE \cite{Forster_Sleight_1999_IntJInorgMat_Y2W3O12NTEExpt,Sumithra_2005}. 

\begin{figure}[t]
\begin{center}
\includegraphics[width=1.0\textwidth]{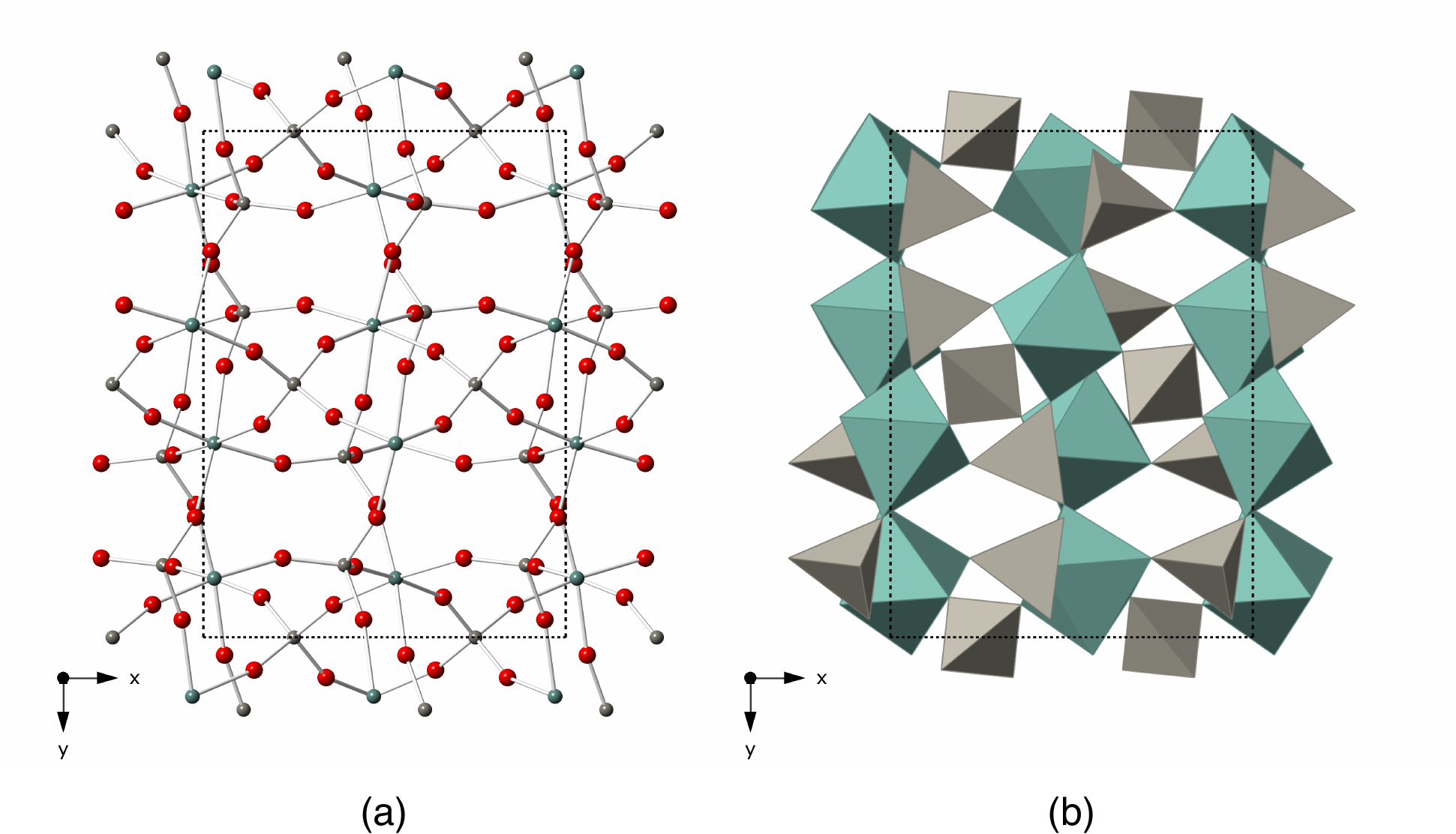} 
\caption[The Y$_2$W$_3$O$_{12}$ structure]{The Y$_2$W$_3$O$_{12}$ structure \cite{Forster_Sleight_1999_IntJInorgMat_Y2W3O12NTEExpt}. (a) Shows the structure in ball-and-stick format: green atoms are Y, grey atoms are W and red atoms are O. (b) Shows the structure in terms of its corner-sharing YO$_6$ and WO$_4$ coordination polyhedra: green octahedra are YO$_6$ units, grey tetrahedra are WO$_4$ units. The unit cell is shown as the dashed black line.}
\label{Y2W3O12picture}       
\end{center}
\end{figure}

\subsection{Y$_2$W$_3$O$_{12}$: an NTE material with no RUMs}

Y$_2$W$_3$O$_{12}$ \cite{Forster_Sleight_1999_IntJInorgMat_Y2W3O12NTEExpt} is a member of the \textbf{A}$_2$\textbf{M}$_3$O$_{12}$ family of NTE materials, where \textbf{A} is a trivalent cation in octahedral coordination, and \textbf{M} is either W or Mo with tetrahedral coordination \cite{Evans_etal_1997_JSolStatChem_NTEinA2M3O12Family,Mary_1999}.  \textbf{A}$_2$\textbf{M}$_3$O$_{12}$ materials have an orthorhombic structure---Y$_2$W$_3$O$_{12}$ is shown in \fref{Y2W3O12picture}---and thus, whilst they can (and in the case of this family, do) have volumetric NTE, their expansion is anisotropic. 

The \textbf{A}$_2$\textbf{M}$_3$O$_{12}$ family is of interest because of its customisability: the \textbf{A} site can withstand a wide range of substitutions, from ions as small as Al$^{3+}$ to ions as large as Ho$^{3+}$ \cite{Lind_2012_Mater_NTEReview}, providing the opportunity to tune thermal expansion behaviour. In general, the smaller the \textbf{A} cation, the smaller the NTE; \emph{in extremis} Al$_3$W$_2$O$_{12}$ actually exhibits weak PTE \cite{Evans_etal_1997_JSolStatChem_NTEinA2M3O12Family}. 

Unlike most \textbf{A}$_2$\textbf{M}$_3$O$_{12}$ compounds, Y$_2$W$_3$O$_{12}$ shows NTE along all three axes \cite{Forster_Sleight_1999_IntJInorgMat_Y2W3O12NTEExpt}. This is unusual for a non-cubic NTE material as they typically  undergo NTE along one or two axes coupled with PTE along the remaining axis or axes \cite{Sleight_1995_Endeavour_NTE}. Furthermore, Y$_2$W$_3$O$_{12}$ has some of the largest coefficients of NTE of the entire \textbf{A}$_2$\textbf{M}$_3$O$_{12}$ family; its average linear expansion coefficient has been measured as $-7.0$~MK$^{-1}$ \cite{Forster_Sleight_1999_IntJInorgMat_Y2W3O12NTEExpt}, which equates to a volumetric coefficient of thermal expansion of $\alpha_{\mathrm{V}}\approx-21$~MK$^{-1}$.  

Y$_2$W$_3$O$_{12}$ is an NTE material of particular interest to us because, if thought of as a series of rigid YO$_6$ and WO$_4$ coordination polyhedra, then the crystal structure would be too over-constrained to support RUMs (this is discussed further in \ref{maxwell_counting}). The existence of NTE in this material therefore suggests a degree of flexibility of either or both of the YO$_6$ octahedra and WO$_4$ tetrahedra, and it is important to understand how this flexibility generates a tension effect of sufficient strength to generate NTE.



Analysis of neutron diffraction data \cite{Forster_Sleight_1999_IntJInorgMat_Y2W3O12NTEExpt} suggests that NTE in Y$_2$W$_3$O$_{12}$ may be driven by transverse vibrations of oxygens within the Y--O--W linkages, as per the tension effect. However, this prediction has yet to be confirmed through analysis of the vibrational spectrum. No experimental study of the phonon spectrum of Y$_2$W$_3$O$_{12}$ has yet been published but parametrised interatomic potential calculations \cite{Sumithra_etal_2007_PRB_Y2W3O12NTEPhononsMechanism} have suggested possible NTE phonons between 1 and 2.5~THz that may correspond to the tension effect. However, this model was only fitted to the structure (sensitive to the first-order differentials of the interatomic potential) and not to any property that will give information about the important second-order differentials (essential for vibrational frequencies). Moreover, the model includes interactions that depend on the Y--O--W bonds but nothing dependent on the more useful O--Y--O or O--W--O bonds.

In this paper we investigate exactly where in the Y$_2$W$_3$O$_{12}$ phonon spectrum the structure flexibility manifests itself and, since the RUM model is not applicable, how NTE phonons exist in enough of the spectrum (and at low enough frequencies) so as to give rise to an overall negative coefficient of thermal expansion. To do this, we use an approach similar to that used in our studies of Zn(CN)$_2$ \cite{Hong_Zn(CN)2}, Cu$_2$O \cite{Leila_Cu2O} and MOF-5 \cite{Leila_MOF5}, that is we perform \textit{ab initio} calculations of the material's phonon spectrum that are then compared to models of structural flexibility. Lattice dynamics calculations have recently been reported for the related material Y$_2$Mo$_2$O$_{12}$ \cite{Wang_Y2Mo3O12,Romao_Y2Mo3O12} which, like Y$_2$W$_3$O$_{12}$, also shows NTE along all three axes \cite{Marinkovic_Y2Mo3O12}. However, the approach taken in this paper (examining the full set of dispersion curves) was not followed in \cite{Wang_Y2Mo3O12,Romao_Y2Mo3O12}.



\section{Methodology}

\subsection{\emph{Ab-initio} calculations}

\begin{table}[t]
\caption[Structural data for the Y$_2$W$_3$O$_{12}$ equilibrium cell]{Structural data for the Y$_2$W$_3$O$_{12}$ equilibrium cell, as optimised at 0~GPa via DFT in CASTEP. Experimental data obtained for Y$_2$W$_3$O$_{12}$ from neutron diffraction at 0~GPa and 15~K \cite{Forster_Sleight_1999_IntJInorgMat_Y2W3O12NTEExpt} are given alongside for comparison. The \emph{ab-initio} results are very close to the experimentally-derived values.}
\begin{center}
\begin{tabular}{| r | c c c | c c c |}
\hline
\hline
 & \multicolumn{3}{c |}{This Study} & \multicolumn{3}{c |}{Experiment} \\
\hline
Space group: & \multicolumn{3}{c |}{62 ($Pnca$)} & \multicolumn{3}{c |}{62 ($Pnca$)} \\
Cell parameters: &\multicolumn{3}{c |}{$a=10.193$ \AA} & \multicolumn{3}{c |}{$a=10.130$ \AA} \\ 
 & \multicolumn{3}{c |}{$b=14.218$ \AA} & \multicolumn{3}{c |}{$b=13.955$ \AA} \\
 & \multicolumn{3}{c |}{$c=10.112$ \AA} & \multicolumn{3}{c |}{$c=10.026$ \AA} \\
\hline
\multicolumn{1}{| c |}{Atom} & $x$ & $y$ & $z$ & $x$ & $y$ & $z$ \\
Y & 0.4697 & 0.3833 & 0.2496 & 0.4669 & 0.3813 & 0.2493 \\ 
W & 0.2500 & 0.0000 & 0.4731 & 0.2500 & 0.0000 & 0.4739 \\ 
W & 0.1112 & 0.3574 & 0.3884 & 0.1167 & 0.3562 & 0.3928 \\ 
O & 0.0839 & 0.1396 & 0.0615 & 0.0905 & 0.1395 & 0.0675 \\ 
O & 0.1402 & 0.0644 & 0.3719 & 0.1343 & 0.0614 & 0.3723 \\ 
O & 0.0217 & 0.2629 & 0.3175 & 0.0177 & 0.2657 & 0.3211 \\ 
O & 0.3387 & 0.4209 & 0.0749 & 0.3332 & 0.4155 & 0.0767 \\ 
O & 0.0583 & 0.4648 & 0.3179 & 0.0732 & 0.4690 & 0.3226 \\ 
O & 0.2812 & 0.3435 & 0.3563 & 0.2868 & 0.3340 & 0.3613 \\ 
\hline
\multicolumn{1}{| c |}{Average distances} & \multicolumn{3}{c |}{} & \multicolumn{3}{c |}{} \\
W--O & \multicolumn{3}{c |}{1.772 \AA} & \multicolumn{3}{c |}{1.775 \AA} \\
Y--O & \multicolumn{3}{c |}{2.267 \AA} & \multicolumn{3}{c |}{2.242 \AA} \\
W--Y & \multicolumn{3}{c |}{3.972 \AA} & \multicolumn{3}{c |}{3.936 \AA} \\
\hline
\end{tabular}
\end{center}
\label{Y2W3O12EquilibriumCell}
\end{table}

Density Functional Theory (DFT) calculations were carried out using the CASTEP code \cite{Clark_etal_2005_ZKrist_CASTEP}. This uses a plane-wave basis set to describe the electronic wave functions together with the pseudopotential method to remove the need to handle core atomic electrons explicitly. We used the GGA-PBE functional \cite{Perdew_etal_1996_PRL_PBEFunctional,Perdew_etal_1997_PRL_PBEFunctionalErrata} and CASTEP's internal on-the-fly-generated pseudopotentials for Y, W and O, regenerated so as to be norm-conserving. The material was explicitly defined as an insulator for all calculations. Forces, stresses and dielectric constants were converged to within 0.005~eV\AA$^{-1}$, 0.01~GPa and 0.00025 respectively. This was achieved using a plane wave cutoff energy of 1200 eV and a $3\times 2\times 3$ Monkhorst-Pack \cite{Monkhorst_Pack_1976_PRB_MPGrids} grid of wave-vectors for the integration of electronic states across the Brillouin zone. Geometry optimisation at 0~GPa gave final structural parameters as detailed in \tref{Y2W3O12EquilibriumCell}. Agreement with experiment is reasonable although, as ever with the GGA model, the calculation slightly overestimates bond distances and lattice parameters. \Tref{Y2W3O12EquilibriumCell} shows that the differences occur mostly in the range $\le1\%$ except for the $b$ lattice parameter (just under  $2\%$ discrepancy).


Phonon frequencies were calculated using Density Functional Perturbation Theory (DFPT) \cite{Baroni_DFPT,Refson_etal_2006_PhysRevB_DFPTPaper} with a $3\times 2\times 3$ Monkhorst-Pack \cite{Monkhorst_Pack_1976_PRB_MPGrids} grid that was offset to place one of the wave-vectors at the origin of reciprocal space. 
Fourier interpolation was used to calculate phonons for wave vectors along high-symmetry directions \cite{Aroyo_etal_2011_BilbaoCrystServer1,Aroyo_etal_2006_ZKrist_BilbaoCrystServer1,Aroyo_etal_2006_ActaCrystA_BilbaoCrystServer2,Tasci_etal_2012_Conference_BilbaoCrystServerKVEC,BilbaoCrystServer} for the production of dispersion curves. In addition, phonons were calculated for 490 wave-vectors spaced randomly throughout the Brillouin zone for the production of the vibrational density of states. This level of sampling of wave vectors proved to be sufficient to achieve convergence of phonon properties.

\subsection{Lattice dynamics calculations and thermal expansion}

The role of individual phonons in thermal expansion is represented by the mode Gr\"{u}neisen parameter $\gamma_{\mathbf{k},i}$ \cite{Barrera_2005_review}, which quantifies the change in phonon frequency with volume through the relation 
\begin{equation}
\gamma_{\mathbf{k},i} = -\frac{\partial \ln \omega_{\mathbf{k},i}}{\partial \ln V}
\end{equation}
where $V$ is the crystal volume and $\omega_{\mathbf{k},i}$ is the angular frequency of the phonon of branch $i$ and wave vector $\mathbf{k}$. 

Within the quasi-harmonic approximation (the approximation in which all temperature dependence of phonon frequencies occurs indirectly through changes in force constants arising from changes in volume) the coefficient of volumetric thermal expansion $\alpha_V$ is given as
\begin{equation}
\alpha_V = \frac{\partial \ln V}{\partial T} = \frac{1}{BV} \sum_{\mathbf{k},i} c_{\mathbf{k},i} \gamma_{\mathbf{k},i}
\end{equation}
where 
\begin{equation}
c_{\mathbf{k},i} = \hbar \omega_{\mathbf{k},i} \frac{\partial n(T,\omega_{\mathbf{k},i})}{\partial T}
\end{equation}
$c_{\mathbf{k},i}$ is the contribution of the individual phonon mode to the overall heat capacity, $T$ is the temperature, $B$ is the bulk modulus, $\hbar$ is the reduced Planck's constant, and $n(T,\omega)$ is the Bose--Einstein distribution for phonons. A negative value of $\gamma_{\mathbf{k},i}$ indicates that the phonon  $i$ at wave vector $\mathbf{k}$ contributes to negative thermal expansion and, if a sufficient number of modes with a large enough negative value of $\gamma_{\mathbf{k},i}$ exist, the overall outcome for the system is a negative value of $\alpha_V$, that is, NTE.

\subsection{Application of lattice dynamics and Gr\"uneisen theory}
Since Y$_2$W$_3$O$_{12}$ has an orthorhombic lattice, changes in cell size ($\Delta V$) were achieved through variable-cell geometry optimisation at an applied hydrostatic pressure. Testing showed that an applied pressure of $-0.06$~GPa  resulted in the desired cell volume change of approximately 0.1\%. The structure was optimised at both this pressure and zero pressure, with phonon frequencies also calculated at both pressures for a set of random wave vectors and for wave vectors along the key symmetry directions in reciprocal space. 

In order to calculate the change in frequency for a particular mode with change in volume, $\Delta \omega_{\mathbf{k},i}$, we used an eigenvector matching algorithm to ensure that we calculated $\omega_{\mathbf{k},i}$ for equivalent modes (\ref{draw}). From these results we generated the set of mode Gr\"{u}neisen parameters numerically as $\gamma_{\mathbf{k},i} =  (\Delta \omega_{\mathbf{k},i} / \omega_{\mathbf{k},i}) / (\Delta V/V)$.

\subsection{Flexibility models}
\label{y2w3o12flexmodelsection}

Our approach was to identify the origin of negative $\gamma_{\mathbf{k},i}$ values in terms of the flexibility of the structure. We therefore constructed a number of flexibility models in which we enforced a high degree of rigidity for certain bonds or structural units, whilst maintaining force constants of zero value for other bond stretching or bending motions. This meant that any phonon calculated for a given flexibility model would have zero or near-zero frequency if that phonon did not violate the constraints of that model, and high frequency otherwise. 

The objective was to compare eigenvectors of phonons from the \textit{ab initio} lattice dynamics calculations with eigenvectors of those zero or near-zero frequency modes calculated for the constructed flexibility models. This allowed us to pinpoint different types of atomic-scale motion in the highly complex \textit{ab initio} dispersion curves (including those regions which drive NTE) with relative ease. 

The flexibility models considered are best described as combinations of separate components that represent different structural units. Hence, the YO$_6$ units were represented by one of
\begin{enumerate}
\item Rigid Y--O rods
\item Rigid YO$_6$ octahedra
\end{enumerate}
\noindent while the WO$_4$ units were represented by one of
\begin{enumerate}
\item Rigid W--O rods
\item Rigid WO$_4$ tetrahedra
\end{enumerate}

\noindent The construction and application of these models is discussed in \ref{flexibility_models}.

It has been suggested that in ZrW$_2$O$_8$, a similar material to Y$_2$W$_3$O$_{12}$ (in that both are frameworks of corner-linked tetrahedral and polyhedral groups of atoms), the NTE phonons may exhibit strong correlations between the motions of adjacent metal sites, with the W--O--Zr bond angles remain undistorted \cite{Cao_etal_2002_PRL_ZrW2O8NTEFrustratedModes,Cao_etal_2003_PRB_ZrW2O8NTEXAFS,Figueiredo_Perottoni_2007_DFTZrW2O8BondStiffness}. 
In order to investigate the possibility of a similar correlation between Y and W atoms in Y$_2$W$_3$O$_{12}$, additional flexibility models were constructed such that Y\ldots W correlations could be taken into account via one of
\begin{enumerate}
\item No Y\ldots W bond
\item Rigid Y\ldots W rod
\end{enumerate}
In total this yielded eight different Y$_2$W$_3$O$_{12}$ flexibility models. However, the two models containing both rigid WO$_4$ tetrahedra and rigid YO$_6$ octahedra are over-constrained (see \ref{maxwell_counting} and \cite{Hammonds_octahedra}) and are, therefore, not discussed here (although models were constructed and tested to confirm this was actually the case).

The six remaining flexibility models were constructed as a set of simple force field models within the GULP lattice simulation program \cite{GULP_2003}. Simple bond stretching and bond angle bending forces were added to each model as discussed in \ref{flexibility_models}, using the ambient pressure structure calculated by CASTEP at 0~GPa.

In order to compare the eigenvectors of the \textit{ab initio} simulation with those of the flexibility model, we define a dimensionless `match' coefficient, $m_{\mathbf{k},i}$, as
\begin{equation}\label{mequation}
m_{\mathbf{k},i} = \Omega^2 \sum_j\frac{\mathbf{e}_{\mathbf{k},i}^\mathrm{phonon}\cdot\mathbf{e}_{\mathbf{k},j}^\mathrm{model}}{\Omega^2+\tilde{\omega}^2_{\mathbf{k},j}} \label{eq:m}
\end{equation}
\noindent where $\mathbf{e}_{\mathbf{k},i}^\mathrm{phonon}$ is the eigenvector of the \textit{ab initio} phonon mode $i$ at wave vector $\mathbf{k}$,  $\mathbf{e}_{\mathbf{k},j}^\mathrm{model}$ is the eigenvector of a mode $j$ in the flexibility model calculated at the same wave vector $\mathbf{k}$, and $\tilde{\omega}_{\mathbf{k},j}$ is the corresponding angular frequency in the flexibility model. $\Omega$ is a scale factor that avoids division-by-zero errors when $\tilde{\omega}_{\mathbf{k},j}=0$, giving instead a value of $m_{\mathbf{k},i}  = \Omega^2/(\Omega^2+\tilde{\omega}^2_{\mathbf{k},j})=1$ when $\tilde{\omega}^2_{\mathbf{k},j}=0$ (we used a value $\Omega = 1$~THz). This sets the range of $m_{\mathbf{k},i}$ values from 0 to 1. A value of $m_{\mathbf{k},i}$ close to 1 occurs when the \textit{ab initio} phonon $i$ is a close match to a flexibility model mode that leaves the defined `rigid' units undistorted. A value of $m_{\mathbf{k},i}$ close to 0 arises when the \textit{ab initio} phonon $i$ involves significant distortions of the rigid units described in the flexibility model. Therefore $m_{\mathbf{k},i}$ is a measure of the extent to which mode $i$ of wave vector $\mathbf{k}$ in the \textit{ab initio} model matches any modes given in the flexibility model at the same wave vector. Our practical implementation of this mode matching is described in \ref{draw} and \ref{flexibility_models}.

\section{Results}

\begin{figure}[t]
\begin{center}
\includegraphics[width=0.7\textwidth]{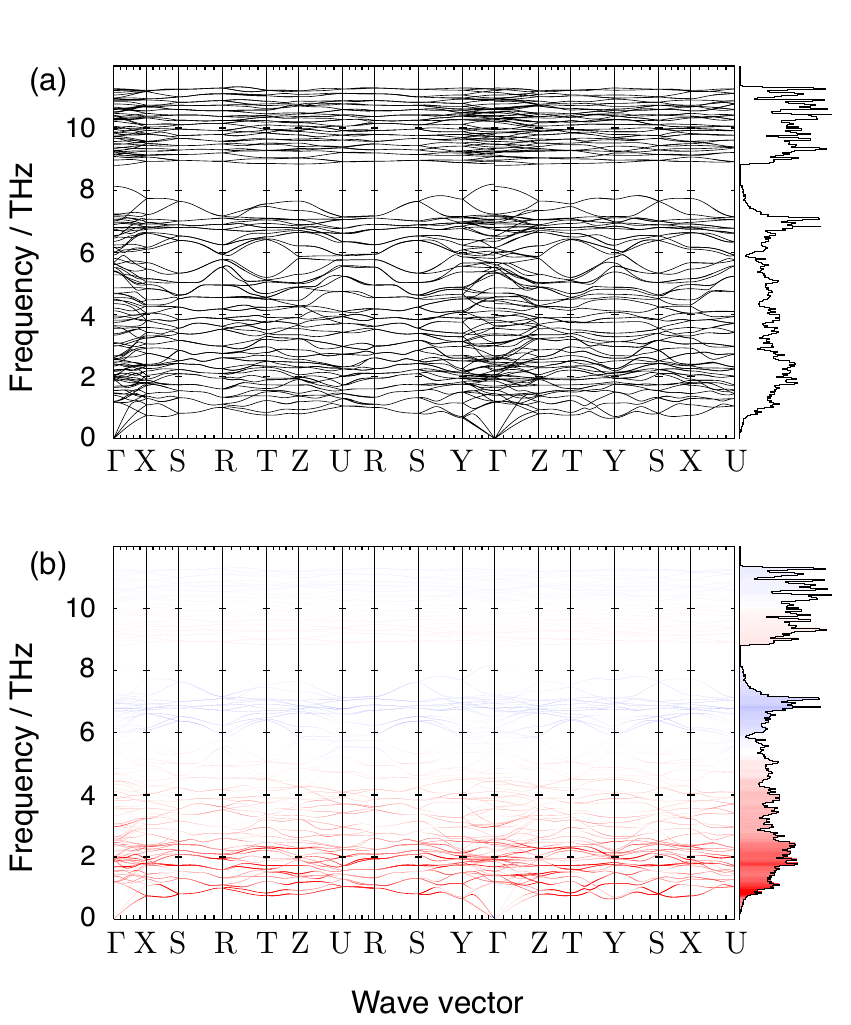} 
\caption[Y$_2$W$_3$O$_{12}$ phonon dispersion curves and densities of states]{(a) Low energy dispersion curves and densities of states for Y$_2$W$_3$O$_{12}$. The notation of Bradley and Cracknell \cite{Bradley_Cracknell_1972_CrystalBZSymmetryBook} is used to denote special Brillouin zone points ($\Gamma \equiv [0,0,0]$, $\mathrm{X} \equiv [\frac{1}{2},0,0]$, $\mathrm{Y} \equiv [0,\frac{1}{2},0]$, $\mathrm{Z} \equiv [0,0,\frac{1}{2}]$). (b) Shows the same data shaded according to the value of $\gamma_{\mathbf{k},i}$ of each mode at each wave vector. The colour scale ranges from red ($\gamma_{\mathbf{k},i}\le-9$) to white ($\gamma_{\mathbf{k},i}=0$) through to blue ($\gamma_{\mathbf{k},i}\ge+9$). Bins that make up the density of states are shaded according to the average $\gamma_{\mathbf{k},i}$ for each bin using the same colour scale.}
\label{Y2W3O12Phonons1} 
\end{center}
\end{figure}

\subsection{Phonon dispersion and mode Gr\"{u}neisen parameters in Y$_2$W$_3$O$_{12}$}
\label{ntephononsiny2w3o12section}


\Fref{Y2W3O12Phonons1}a shows the calculated phonon dispersion curves and density of states of Y$_2$W$_3$O$_{12}$ in the 0--12~THz frequency range. As can be seen, Y$_2$W$_3$O$_{12}$ dispersion curves are relatively complex due to many mode crossings and the existence of multiple phonons within a tight frequency range. Nevertheless, it can be seen that the density of states is a continuous function up to about 8~THz, with a gap before another band of phonons in the range 9--11.5~THz. Not shown in the diagram is a further tight band of higher-frequency vibrations around 24~THz and a final set of vibrations between 27--30~THz.

There are 204 separate modes $i$ for any wave vector $\mathbf{k}$. The two highest-frequency bands (24 and 27--30~THz) consist of 48 individual modes, the band between 9--11.5~THz consists of 60 individual modes, and the broad band from 0--8~THz consists of the other 96 modes. We discuss eigenvectors associated with these bands in the next subsection.

\Fref{Y2W3O12Phonons1}b shows the dispersion curves with colours used to represent the values of the associated mode Gr\"{u}neisen parameters $\gamma_{\mathbf{k},i}$ (red for negative and blue for positive values respectively). It is interesting to note that the dispersion diagram is divided into large bands of negative or positive values of mode Gr\"{u}neisen parameters. This feature stands in contrast to other recently studied systems such as Cu$_2$O \cite{Leila_Cu2O}, ScF$_3$ \cite{ScF3_phonons}, Zn(CN)$_2$ \cite{Hong_Zn(CN)2} and MOF-5 \cite{Leila_MOF5}, where phonons with significantly different values of mode Gr\"uneisen parameters can exist at similar frequencies due to the significant dependence of the NTE phonon eigenvectors on wave vector. The strongest contributions to NTE come from phonons around 0.7--1~THz ($\gamma_{\mathbf{k},i}\sim-9$ on average, with values of $\sim-15$ at wave vectors near the Brillouin zone points S and T). Acoustic modes also contribute to NTE, as is the case in some other systems recently studied \cite{Leila_Cu2O,Hong_Zn(CN)2,Leila_MOF5}. We note that recent \textit{ab initio} calculations of the mode Gr\"uneisen parameters at the $\Gamma$ point in Y$_2$Mo$_3$O$_{12}$ using the frozen-phonon method also show a large number of modes with negative values, but see smaller values except for the lowest-frequency mode \cite{Wang_Y2Mo3O12}.
 
A second group of strongly-NTE phonons exists around 1.8~THz but, more generally, it can be seen that all the phonons up to around 5~THz have negative $\gamma_{\mathbf{k},i}$ (albeit approaching zero values as the frequency increases to 5~THz). There is also a higher frequency band of weak NTE modes around 9--10~THz $\gamma_{\mathbf{k},i}\sim0$. Weak, low frequency, PTE modes exist around 5--8~THz and for frequencies larger than 10~THz.
 


\subsection{Flexibility model mapping}

\begin{figure}[t]
\begin{center}
\includegraphics[width=\textwidth]{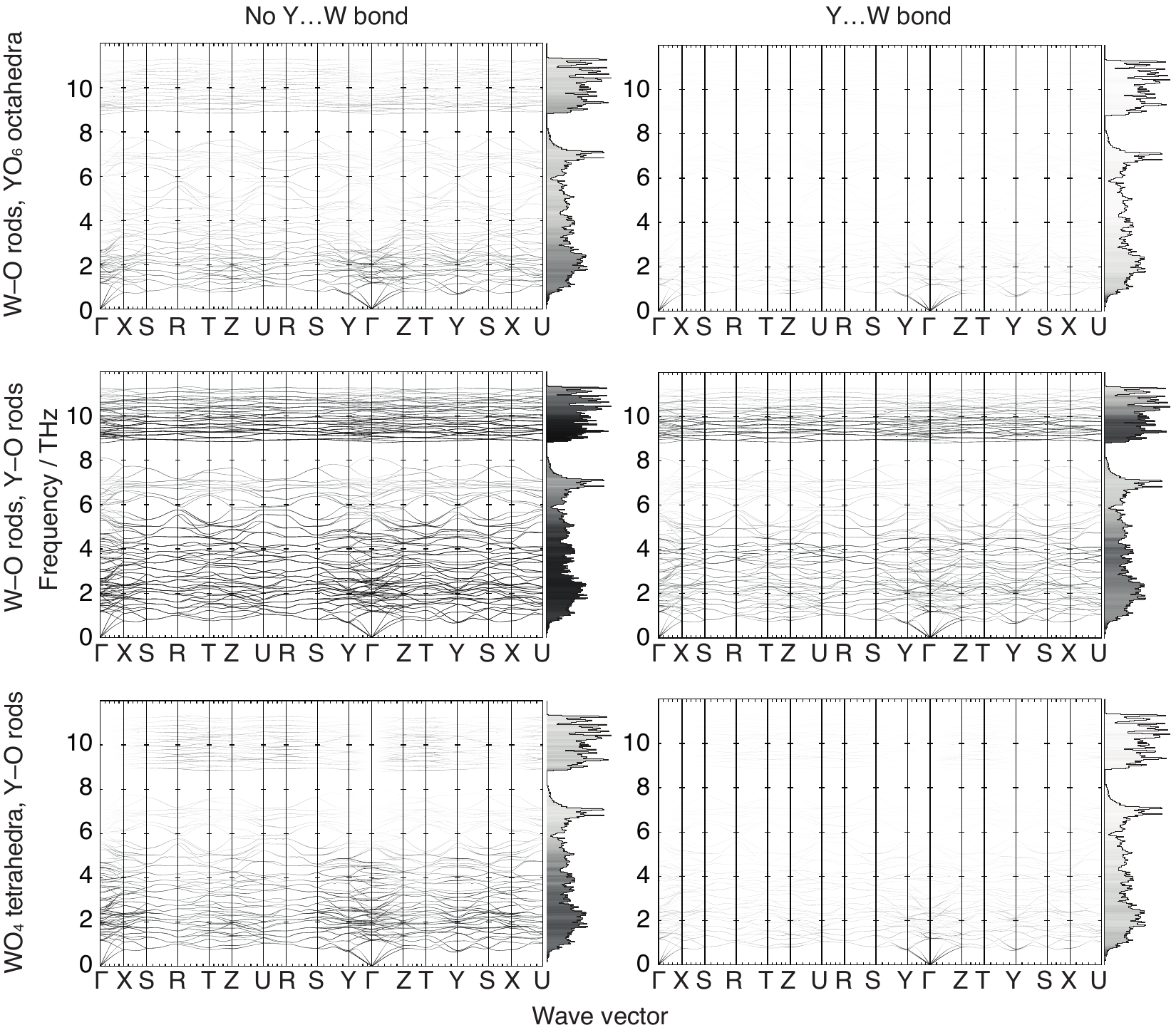} 
\caption[Flexibility analysis of Y$_2$W$_3$O$_{12}$ phonon dispersion curves]{Flexibility analysis of Y$_2$W$_3$O$_{12}$ phonon dispersion curves and densities of states. All flexibility models shown here describe YO$_6$ units in terms of corner-sharing rigid YO$_6$ octahedra. The notation of Bradley and Cracknell denotes special Brillouin zone points \cite{Bradley_Cracknell_1972_CrystalBZSymmetryBook}. All data are shaded according to the value of $m_{\mathbf{k},i}$ at each mode for each wave vector. The shading ranges from white ($m_{\mathbf{k},i}=0$) through to black ($m_{\mathbf{k},i}=1$). Bins that make up the density of states are shaded according to the average $m_{\mathbf{k},i}$ for each bin using the same colour scale.}
\label{Y2W3O12Flexibility} 
\end{center}
\end{figure}

\Fref{Y2W3O12Flexibility} shows six plots of the 156 Y$_2$W$_3$O$_{12}$ \textit{ab initio} dispersion curves for frequencies between 0--12~THz (as in \fref{Y2W3O12Phonons1}) together with their corresponding densities of states. In each case the  data are shaded in greyscale according to the degree to which one of the six flexibility models can reproduce the eigenvectors of each phonon in the \textit{ab initio} calculations. Black shading indicates a perfect match between the \textit{ab inito} phonons and the flexibility model in question; white shading indicates no relation between the two. 
This approach offers a convenient, visual sense of the atomic motions associated with different regions of the complex Y$_2$W$_3$O$_{12}$ phonon spectrum. Further details on how these `match' values were calculated are given in \ref{flexibility_models}. 

In what follows we use the information presented in \fref{Y2W3O12Phonons1} and \fref{Y2W3O12Flexibility} to correlate the NTE phonons identified in \sref{ntephononsiny2w3o12section} with the vibrations permitted by the six flexibility models.


\subsubsection{Rigid W--O and Y--O rods with no rigid polyhedra.}

As discussed in \ref{maxwell_counting}, a flexibility model of Y$_2$W$_3$O$_{12}$ consisting of rigid W--O and Y--O rods has 96 excess degrees of freedom. We would therefore expect many of the 156 lower-frequency Y$_2$W$_3$O$_{12}$ phonons to correspond to rigid W--O and Y--O rod motions. \Fref{Y2W3O12Flexibility} shows that this is largely the case. Much of the \textit{ab inito} phonon spectrum in the 0--12~THz range is a strong match for rigid W--O/Y--O flexibility model modes. However, this motion is not confined to specific phonons in the spectrum but, instead, there is a high degree of eigenvector mixing and the flexibility of the model is spread over multiple phonons in the real material. This mixing follows from the relatively low symmetry and the complexity of the dispersion curves.

Despite the high degree of eigenvector mixing present, it is still possible to observe general trends within the phonon spectrum. Specifically, phonons in the 0--5~THz (approximately 64 modes) and 9--10~THz (approximately 28 modes) ranges have the strongest match for rigid W--O and Y--O motions. These differences correspond well with the separation of PTE and NTE phonons in the Y$_2$W$_3$O$_{12}$ spectrum as seen in \fref{Y2W3O12Phonons1}. NTE phonons have the strongest match with the flexibility model (corresponding to negligible W--O and Y--O bond stretching) while PTE phonons have the weakest match (corresponding to small, but non-trivial, stretching of W--O and Y--O bonds). This observation is a good quantification of the tension effect giving rise to NTE in Y$_2$W$_3$O$_{12}$.

The addition of the constraint that keeps the Y\dots W distance fixed yields similar results. This flexibility model has 60 excess degrees of freedom (reduced from the original 96; see \ref{maxwell_counting}). \Fref{Y2W3O12Flexibility} shows that the same eigenvector mixing that was observed previously is still present, and that the reduced flexibility of this model manifests itself in a reduction in the match with the \textit{ab initio} phonons (seen in the lighter greyscale shading in \fref{Y2W3O12Flexibility}) across the whole phonon spectrum. We can conclude that the eigenvectors of all phonons with frequencies in the range 0--12~THz in Y$_2$W$_3$O$_{12}$ involve some changes in Y\dots W distance.

\subsubsection{Rigid WO$_4$ tetrahedra and Y--O bonds.}

The analysis of \ref{maxwell_counting} shows that the flexibility model for Y$_2$W$_3$O$_{12}$ consisting of rigid WO$_4$ tetrahedra and Y--O rods has 48 excess degrees of freedom. This model is still highly flexible, albeit less flexible than the two rigid rod models discussed above.

\Fref{Y2W3O12Flexibility} shows, once again, that a high degree of eigenvector mixing occurs for this system with each flexibility model phonon spread across multiple \textit{ab initio} phonons. It can also be seen that the lower overall flexibility manifests itself in the lightening of the greyscale, indicating a weaker overall match with the \textit{ab inito} phonons. This lightening is not uniform; match values are still relatively strong in the 0--5~THz region and approach a perfect match for the lowest frequency phonons. Above 5~THz the match between this model and the \textit{ab initio} phonons decreases with increasing frequency. The largest reduction in the match with the flexibility model is in the 9--11.5~THz region. We can identify this region with the modes involving significant bending of the bonds within the WO$_4$ tetrahedra. The analysis of \ref{maxwell_counting} shows that the number of WO$_4$ bond-bending modes is 60, which is exactly the number of modes in the 9--11.5~THz region.

The phonons in the 0--5~THz range also correspond to the strongest NTE-driving phonons. This, together with the above observations, suggests that these lowest frequency NTE modes correspond to motion of near-rigid WO$_4$ tetrahedra and rigid Y--O rods. 

The addition the Y\dots W distance constraint leads to a very weak match between the flexibility model phonons and those of the \textit{ab initio} calculation. This can be seen in the considerable lightening of the greyscale shading for this model in \fref{Y2W3O12Flexibility}. It is consistent with the analysis of \ref{maxwell_counting}, which shows that adding the additional Y\dots W distance constraint greatly reduces the flexibility of this model from 48 to exactly zero excess degrees of freedom. In short, a model with rigid WO$_4$ tetrahedra is unable to support motions in which the Y\dots W distance is  conserved.

\subsubsection{Rigid YO$_6$ octahedra and W--O bonds}

It is seen from \ref{maxwell_counting} that a flexibility model consisting of rigid YO$_6$ octahedra and W--O rods has 36 excess degrees of freedom, which is fewer than for the rigid WO$_4$ tetrahedron/Y--O rod model discussed above. Accordingly, \fref{Y2W3O12Flexibility} shows that the overall match between this flexibility model and the \textit{ab initio} dispersion curves is lower than for the rigid WO$_4$/Y--O model. Once again, a significant amount of eigenvector mixing is also apparent.

The strength of the match between this model and the \textit{ab initio} phonons is greatest for the lowest frequency NTE phonons in the 0--2~THz range. In general, the match is weaker than that found for the rigid WO$_4$ tetrahedra/Y--O rod model and, aside from the acoustic phonons at the $\Gamma$ point, a perfect match is never achieved. There is one exception to this; in the 9--10~THz range the match between this model and the \textit{ab initio} phonons \textit{increases} slightly and is slightly stronger than that for the rigid WO$_4$ tetrahedra/Y--O rod model, which is consistent with the identification of the frequency region above 9~THz corresponding to WO$_4$ bond-bending modes. At frequencies above 10~THz the match values decrease once again.


This all suggests that the full phonon spectrum of Y$_2$W$_3$O$_{12}$ involves some degree of O--Y--O bond angle bending. The amount of bending is least in two NTE regions of the phonon spectrum: the strongest NTE phonons in the 0--5~THz frequency band and the weak NTE phonons in the 9--10~THz frequency band. The latter is also a region where significant O--W--O bond angle bending can been seen. Otherwise, the general trend observed is for the degree of O--Y--O bond angle bending to increase as a function of increasing frequency.


The analysis of \ref{maxwell_counting} shows that the addition of the Y\dots W distance constraint leads to the flexibility model having more constraints than degrees of freedom. This result is confirmed by the corresponding data plotted in \fref{Y2W3O12Flexibility}, which show a negligible match between the flexibility model and the \textit{ab initio} phonons.

\section{Discussion}

In materials where NTE has been studied through lattice dynamics calculations of dispersion curves and associated mode Gr\"{u}neisen parameters (for example, Cu$_2$ \cite{Cu2O_Gupta,Leila_Cu2O}, ScF$_3$ \cite{ScF3_phonons}, Zn(CN)$_2$ \cite{Hong_Zn(CN)2}, MOF-5 \cite{Leila_MOF5}), it has usually been found that there is a well-defined set of low-frequency modes with negative mode Gr\"uneisen parameters. In this study we have found that Y$_2$W$_3$O$_{12}$ is unusual in that a high degree of eigenvector mixing occurs, leading to the existence of broad bands of phonons with either negative or positive mode Gr\"{u}neisen parameters. 

More specifically, this study has found that NTE in Y$_2$W$_3$O$_{12}$ is primarily driven by phonons in the 0--5~THz frequency range, with the strongest contribution coming from those phonons at the lowest frequencies. There is also some weaker contribution to NTE from another band of phonons in the 9--10~THz frequency range. All other phonons drive PTE.

We had previously shown that any material containing octahedra within a corner-linked network with no non-bridging bonds cannot support RUMs \cite{Hammonds_octahedra} (the ScF$_3$ structure being the one exception, as discussed in reference \cite{Dove_RUMs}). Thus it follows that NTE in Y$_2$W$_3$O$_{12}$ does not arise from the `traditional' RUM mechanism \cite{NTE_theory_Welche} based on rigid YO$_6$ octahedra and WO$_4$ tetrahedra. Our flexibility analysis reported here has found that the NTE phonons in this system involve motion of rigid Y--O and W--O rods as per the tension effect enabled by relatively stiff cation-oxygen bonds. 

However, the tension effect appears to act here as part of a more complex process. In the 0--5~THz range (which contributes strongly to NTE) the WO$_4$ tetrahedra remain effectively rigid with minimal O--W--O bond angle bending while the YO$_6$ octahedra undergo significant O--Y--O bond angle deformation. In the 9--10~THz range (which contributes less to NTE) both O--W--O and O--Y--O bond angles undergo non-trivial deformation although the O--Y--O angles actually deform less than the O--W--O. All PTE regions involve non-trivial levels of Y--O and W--O bond stretching.

As noted earlier, the RUM model is only applicable to systems with stiff coordination polyhedra and is therefore not applicable to Y$_2$W$_3$O$_{12}$. The results of this study show that, in Y$_2$W$_3$O$_{12}$, such very low frequency deformations are made possible by the relatively small energy cost of O--Y--O bond angle bending in YO$_6$ units. 
The combination of the flexibility of the O--Y--O bond angle and the existence of effectively-rigid WO$_4$ tetrahedra and Y--O rods is such that macroscopic NTE is still achieved. By contrast, the low flexibility of the O--W--O bond angle forces the second band of NTE phonons (which resemble the `traditional' tension effect) to a much higher frequency range of 9--10~THz and, thus, they cannot drive macroscopic NTE on their own.



One issue that remains unexplained is the variation in the size of the negative mode Gr\"uneisen parameter for the lowest frequency NTE phonon at the S and T points in reciprocal space. The flexibility models considered here do not highlight anything unusual happening at these points; nevertheless Gr\"uneisen parameters do become more negative, changing from $\sim-9$ to $\sim-15$ as these points are approached.

\section{Summary}

We have investigated the origins of negative thermal expansion in Y$_2$W$_3$O$_{12}$, a structure that is too over-constrained to support RUMs but does, nevertheless, exhibit strong NTE. \emph{Ab-initio} lattice dynamics calculations were used to generate phonon data that were analysed using a variety of flexibility models that identified NTE as being driven primarily by vibrations of effectively rigid WO$_4$ tetrahedra and  Y--O rods. Such a mechanism can exist at the low frequencies necessary for macroscopic NTE because deformation of the O--Y--O bond angle in Y$_2$W$_3$O$_{12}$ involves very little energy cost.


We have also observed that, unlike other recently-studied NTE materials, there is a high degree of eigenvector mixing occurring in Y$_2$W$_3$O$_{12}$. Eigenvectors associated with motions such as bond-bending are mixed across phonons spanning the whole frequency range so completely that there are no modes significantly associated with just one type of motion. As a result, unlike many NTE systems where specific phonons drive NTE, in Y$_2$W$_3$O$_{12}$ NTE is characterised by large bands of phonons with the same sign---and similar value---of mode Gr\"{u}neisen parameter.

\section*{Acknowledgements}
The authors are grateful to Keith Refson for his helpful advice and input. We are grateful for support from NERC (NE/I528277/1, studentship for LHNR),  Innovate UK (KTP009358), and CrystalMaker Software Ltd. Via our membership of the UK's HPC Materials Chemistry Consortium, which is funded by EPSRC (EP/F067496), this work made use of the facilities of HECToR, the UK's national high-performance computing service, which is provided by UoE HPCx Ltd at the University of Edinburgh, Cray Inc and NAG Ltd, and funded by the Office of Science and Technology through EPSRC's High End Computing Programme.

\appendix

\section{Flexibility analysis}\label{maxwell_counting}

At several points in this paper we use a rigidity analysis following the approach originally developed by James Clerk Maxwell \cite{Maxwell_1864}. In a `Maxwell count' we may consider a structure to be a set of points (in this case atoms) connected by rods (in this case bonds). Each point has three degrees of freedom ($x$, $y$ and $z$ translations) and each rod is taken to be rigid and thus represents a constraint. The  number of excess degrees of freedom $N_\mathrm{e}$ is determined by the number of degrees of freedom $N_\mathrm{f}$ and the number of constraints $N_\mathrm{c}$ such that,
\begin{equation}
 N_\mathrm{e} = N_\mathrm{f} - N_\mathrm{c}
\end{equation}
\noindent The larger the positive value of $N_\mathrm{e}$, the more flexible the structure. Conversely, if $N_\mathrm{e}$ is negative, then the structure is over-constrained and has no flexibiility.


To illustrate this approach we can calculate the constraints and degrees of freedom of an isolated WO$_4$ tetrahedron. This has 5 atoms and therefore $N_\mathrm{f}=5\times3=15$ degrees of freedom, while the 4 W--O bonds contribute 4 constraints. To make the tetrahedron rigid requires an additional 5 constraints on the internal bond angles (these can be thought of in terms of adding O--O bonds but, whilst there are 6 such bonds in a WO$_4$ tetrahedron, only 5 are required to make the structure rigid with the sixth being redundant). $N_\mathrm{c} = 4 + 5 = 9$ and thus, the number of excess degrees of freedom of a rigid WO$_4$ tetrahedron is $N_\mathrm{e}=15-9 =6$, which corresponds to the number of degrees of freedom of a three-dimensional rigid body (i.e. 3 translational plus 3 rotational).

Similarly, for an isolated YO$_6$ octahedron with 7 atoms, $N_\mathrm{f}=7\times3=21$. The octahedron also contains 6 Y--O bonds which act as constraints and, to make the octahedron completely rigid, it requires another 9 constraints on its internal bond angles. Thus $N_\mathrm{c} = 6 + 9 = 15$ and $N_\mathrm{e}=21-15=6$. Once again, this is the standard result for a three-dimensional rigid body.

Applying Maxwell counting to the crystal structure of Y$_2$W$_3$O$_{12}$, we note that the system has 4 formula units per unit cell with Y--O--W linkages between WO$_4$ tetrahedra and YO$_6$ octahedra and no non-bridging bonds. Modelling Y$_2$W$_3$O$_{12}$ as a series of unconstrained atoms, we have $N_\mathrm{f} = 17 \times 3 \times 4 = 204$; this number corresponds to the number of normal modes for any wave vector for this crystal. 

Using the above, we can now add the constraints associated with the flexibility models of Y$_2$W$_3$O$_{12}$ introduced in \sref{y2w3o12flexmodelsection}. First, we consider the model with rigid Y--O and W--O rods but otherwise flexible polyhedra. A single unit cell contains 8 YO$_6$ octahedra and 12 WO$_4$ tetrahedra and, thus, 48 Y--O bonds and 48 W--O bonds. For this model, $N_\mathrm{c}=48+48=96$ and $N_\mathrm{e}=204-96 =108$. The model comprising rigid Y--O rods and WO$_4$ tetrahedra has 48 constraints from the rods plus $9 \times12=108$ constraints from the tetrahedra, giving $N_\mathrm{c}=108+48=156$ and $N_\mathrm{e} = 204-156=48$. The model comprising rigid W--O rods and YO$_6$ octahedra has 48 constraints from the rods plus $15 \times 8=120$ constraints from the octahedra, giving $N_\mathrm{c} = 120+48=168$ and $N_\mathrm{e} = 204-168=36$. Meanwhile, a `classic' RUM model comprising rigid WO$_4$ tetrahedra and YO$_6$ octahedra has 108 constraints from the tetrahedra and 120 constraints from the octahedra, giving $N_\mathrm{c} = 108+120=228$ and $N_\mathrm{e} = 204-228=-24$. This result is consistent with the assertion that a system of corner-linked rigid tetrahedra and octahedra with no non-bridging bonds is over-constrained and cannot support any RUMs \cite{Hammonds_octahedra}.

An additional constraint, in the form of a Y--W rod, can be incorporated to explore the possibility of the Y\dots W distance in Y$_2$W$_3$O$_{12}$ remaining unchanged. Each unit cell would contain 48 Y\dots W rods, therefore, if we consider an otherwise unconstrained system containing only 8 Y and 12 W atoms then $N_\mathrm{f} = (8+12)\times3=60$. The addition of Y\dots W rods gives $N_\mathrm{e} = 60-48=12$. Such a system can, therefore, support motion in which the structure flexes without changing any of the Y\dots W distances.

Applying a Y\dots W rod to the flexibility models described above causes the number of excess degrees of freedom to fall to 60  for the Y--O/W--O rod model; 0 for the WO$_4$ tetrahedron/Y--O rod model and $-12$ for the YO$_6$ octahedron/W--O rod model. Thus, the rigid rod model remains flexible to some extent with a rigid Y\dots W distance; the WO$_4$ tetrahedron/Y--O rod model reaches the balance between flexibility and rigidity, and the YO$_6$ octahedra/W--O rod model becomes over-constrained.


\section{Plotting enhanced dispersion curves}\label{draw}

All results in this paper were processed using a program written specifically for this work. To draw dispersion curves, the program takes data in the form of phonon frequencies and eigenvectors for a given set of wave vectors. It uses these eigenvectors to match modes from one wave vector to the next via a calculation of $\mathbf{e}_\mathbf{k} \cdot \mathbf{e}_{\mathbf{k} + \delta \mathbf{k}}$, where $\mathbf{e}_\mathbf{k}$ is any phonon eigenvector at wave vector $\mathbf{k}$ and neighbouring wave vectors in the calculation differ by $\delta \mathbf{k}$. For small $\delta \mathbf{k}$, only one of the products of eigenvectors will have a value close to unity, and this continuity of points between wave vectors is easy identified. 

Dispersion curves are drawn directly and saved as an encapsulated PostScript (EPS) file, rather than by printing out a list of ordered data for another plotting program. This allows for the plotting of dispersion curves for many directions in reciprocal space simultaneously. \Fref{Y2W3O12Phonons1}a is an example of this application.

The same eigenvector matching approach is used when numerically evaluating the difference in frequencies between identical modes for lattice dynamics calculations from different volumes. Here, the program takes as input two sets of phonon frequencies and their eigenvectors. Given phonon data calculated for the same material and the same wave vectors but at different pressures/volumes, the program initially matches up equivalent modes for the two different pressures/volumes via the comparison of eigenvectors using the product $\mathbf{e}_\mathbf{k} \cdot \mathbf{e}^\prime_{\mathbf{k}}$, where $\mathbf{e}_\mathbf{k}$ and $\mathbf{e}^\prime_{\mathbf{k}}$ are eigenvectors of phonons at the same wave vector but different volumes. It then calculates the mode Gr\"{u}neisen parameters using the two sets of mode frequencies. When dispersion curves are plotted using the above approach, each mode is shaded according to its corresponding value of the mode Gr\"{u}neisen parameter: red is chosen for negative values, blue for positive values, with both colours graduating towards white as values approach zero. \Fref{Y2W3O12Phonons1}b is an example of this application.

The third capability of the program is to compare the eigenvectors of two separate phonon calculations, calculating a `match' parameter using \eref{mequation} and converting this match value to a grey-scale value for the small section of the plotted dispersion curve, with black representing a value of 1 and white representing a value of 0. \Fref{Y2W3O12Flexibility} is an example of this application.

The program was written in a modern dialect of Fortran (2008). It is available from the authors of this paper (send email to the corresponding author), together with a detailed instruction document. Other examples of its use are to be found in references \cite{Leila_Cu2O}, \cite{Hong_Zn(CN)2} and \cite{Leila_MOF5}.

\section{Implementation of flexibility models}\label{flexibility_models}
As in our work on Cu$_2$O \cite{Leila_Cu2O}, we constructed simple models that represent an idealised flexibility of the Y$_2$W$_3$O$_{12}$ crystal structure. Bonds included in any flexibility model were simulated using a potential energy function of the form
\begin{equation} \label{bond_stretch_energy}
E_\mathrm{bond} = \frac{1}{2} k (r-r_0)^2
 \end{equation}
where $E_\mathrm{bond}$ is the energy of the bond, $r$ is the length of a bond, $r_0$ represents the expected bond length, and k is a constant with a large value. Bond angles were represented by a similarly simple function of the form
\begin{equation} \label{angle_stretch_energy}
E_\mathrm{angle} = \frac{1}{2} K (\theta-\theta_0)^2
 \end{equation}
where $E_\mathrm{angle}$ is the energy of the bond, $\theta$ is the angle of a bond,  $\theta_0$ is an `ideal' (typically of order $90^\circ$ or $109.47^\circ$), and $K$ is a constant.

The intention of this model is to give zero or near-zero (relative to the value of $\Omega$ in \eref{mequation}) frequencies for phonons that represent flexible modes of the model; namely modes whose eigenvectors do not involve stretching or bending of the specified bonds. The exact values of $k$ and $K$ used in \eref{bond_stretch_energy} and \eref{angle_stretch_energy} will not change the eigenvectors of these modes (which are fixed by geometry and topology) and will primarily change the frequencies only of the modes that involve stretching and bending of bonds and whose frequencies will be larger than $\Omega$. Tuning of these parameter values against frequencies of the real material is not desirable, since the models are only abstractions designed to identify the rigid and flexible regions of the crystal structure and are not intended reflect any other properties of the real system. The flexibility models used in this paper were implemented using the GULP lattice modelling program \cite{GULP_2003}.


\end{document}